\documentstyle[12pt]{article}

\newcommand{\bb}{\begin{equation}}
\newcommand{\ee}{\end{equation}}
\newcommand{\ba}{\begin{array}}
\newcommand{\ea}{\end{array}}
\newcommand{\beqa}{\begin{eqnarray}}
\newcommand{\eeqa}{\end{eqnarray}}
\newcommand{\MS}{$\bar{MS}$ }

\begin{document}

\title{Reconciliation of effective potential calculations of the lower
       bound for the Higgs boson mass}
\author{R.S.Willey\thanks{\it Physics Dept, University of Pittsburgh,
Pittsburgh PA 15260;    E-Mail:willey@vms.cis.pitt.edu}}

\date{December 11,1995}

\maketitle

    There are three recent calculations of the lower bound for the mass of
the Higgs boson of the Standard Model, based on the position of the
minimum of the effective potential. \cite{SAC} (I will refer to these as
S,AI,CEQ) In all cases, the result depends on $m_t,\alpha_s$, and on the
cutoff scale beyond which new physics is assumed to set in. For a high
scale ($\sim m_{PL}$) cutoff, the results are in the range $130$ to $140$
GeV, for $m_t = 174, \alpha_s = .118$.  Five to ten GeV is also the claimed
  uncertainty in the high scale case.
However, for the low scale cutoff ($\sim$ one TeV), there is a substantial
discrepancy. AI obtain for the lower bound $72$ GeV; CEQ obtain $55$ GeV. (
S does not give a low scale result for these values of $m_t,\alpha_s$).
\cite{W} The claimed uncertainty in the low scale calculations is only a
few GeV. If this discrepancy persists, the reliability of the general
procedure is placed in doubt.

    I believe that I have traced the origin of the discrepancy to the
treatment of the relation between the perturbative pole mass and the
the\MS mass of the top quark in CEQ. Equation ($18$) of CEQ is the
  relation between these
quantities in QCD; but when the Higgs-Yukawa sector is included there
are additional contributions  which are larger than the QCD
contribution and of the opposite sign.\cite{BWHK}
\bb
 \ba{c}
   m^*=m\{1+\frac{\alpha_s}{\pi}C_F +\frac{y^2}{16 pi^2}\frac{1}{2}\Delta (r)
 +\delta \overline{\zeta}_v \}  \medskip  \\
\delta \overline{\zeta}_v = \frac{1}{16 \pi^2}[3\lambda
(1-\ln{\frac{M^2}{m^2}}) -2 N_c y^2 \frac{m^2}{M^2} +\frac{3}{2}g_2^2
\frac{M_W^2}{m^2}(1-
\ln{\frac{M_W^2}{m^2}}) + \frac{3}{4}\frac{g_2^2}{c^2}\frac{M_Z^2}{m^2}(1-
\ln{\frac{M_Z^2}{m^2}})]  \medskip  \\
  \Delta (r) = -\frac{1}{2}+\int^1_0 dx (2-x)\ln{((1-x)r^2 +x^2)}
 \ea
 \label{1}
\ee
In these equations $m^*$ is the top pole mass, m is the \MS top mass,and M is
the Higgs mass. r is the ratio $\frac{M}{m}$ and y is the top Yukawa
coupling ($y=\sqrt{2}\frac{m}{v}$). $\delta \overline{\zeta}_v$ is the
finite part of the one-loop tadpole contribution which is not removed by
\MS subtraction. \cite{BWHK}(There are also small contributions proportional
 to the gauge coupling constants $g_2^2,g_1^2$ which are neglected in the first
line) Note the term proportional to $\frac{m^2}{M^2}$.
One sees immediately that there will be large corrections for the low scale
(lighter Higgs) case and much smaller correction for the high scale case,
which is what is required.

    CEQ use the QCD relation to fix the  values $m(m^*),y(m^*)$
 for the \MS top mass and Yukawa coupling constant, $m(t),y(t)$, for
a given input top mass e.g. $m^* = 174$. The actual calculation of CEQ,
given this input, is rather complicated, but we can estimate the effect of
using (\ref{1}) in place of CEQ eq (18) to make this connection, as follows:
First, for some input $m^*$, solve (\ref{1}) for $m(m^*)$.Then solve CEQ
eq (18) for a new $m^{*'}$ required to produce that same m. Then the estimate
for the corrected lower bound Higgs mass is the value read from fig 5 or
fig 6 of CEQ for $m^{*'}$.
 There is one new feature. The more complicated (\ref{1})
depends on the Higgs mass M. So one has to guess the output M and put it into
(\ref{1}) and do  the calculation, and then iterate if neccessary. For
the low scale cutoff calculation, we put $M=77$ into (\ref{1}) and find
 $m(174)=198$. We substitute that value into eq (18) of CEQ to
find what $m^{*'}$ CEQ would have to use to get that initial value. The
answer is $m^{*'} = 207$. By extrapolation of fig 5 of CEQ to $m^{*'}=207$,
the corrected lower bound for the Higgs mass looks to be close to 77 GeV.
For the high scale calculation, we start by guessing $M=150$ to put in
(\ref{1}). Now the QCD and nonQCD terms almost exactly cancel,
 m(174) = 174. Then $m^{*'}=182$. For this value of $m^{*'}$,
fig 6 of CEQ gives the lower bound to be 150 GeV. This is to be compared to
values 141 GeV from S and 135 Gev from AI.

    These numbers are not to be taken entirely seriously. They are the
results of a simple patch job, not a real calculation. (There is also the
problem of reading data from photographically reduced figures). However I
do take them as strong indication that the real results of the CEQ
calculation may be consistent with those of S and AI, with uncertainties
which are only a little larger than claimed (and are larger for the large
extrapolation involved in the high scale case than in the low scale limitt).


\begin{thebibliography}{10}
\bibitem{W} I have recently given an alternative derivation of a lower
bound for the Higgs mass which does not make use of the effective
potential. For the same input parameters the results are 72 GeV for
the low scale cutoff and 148 GeV for the high scale cutoff. hep-ph9512226.
\end{thebibliography}
\end{document}